%% file: main.tex
\documentclass[epj]{svjour}
\usepackage{graphics,rotating}
\usepackage{epsfig}
\begin{document}
\title{Effect of spin fluctuations on T$_{c}$ from density-functional theory for superconductors.}
\author{Ma\l{}gorzata Wierzbowska}
\institute{INFM DEMOCRITOS National Simulation Center, via Beirut 2--4, 34014 Trieste, Italy}
\date{Received: date / Revised version: date}
%
\abstract{
The transverse spin fluctuations are introduced to
the density functional theory for superconductors (SCDFT).
Paramagnons are treated within  the random phase approximation
and assumed to be the same for the normal and superconducting state.
The effect of spin fluctuations on T$_{c}$ is studied for a few
simple metals at ambient pressure and niobium at several pressures up to 80~GPa.
} 
\maketitle



\input{introduction}

\input{gap}

\input{self}

\input{normal}

\input{magnetic}

\input{implementation}

\input{results}

\input{summary}


\input{refs}
\end{document}

%% file: introduction.tex
\section{Introduction}

Since the discovery of superconductivity many theories have been born to explain
this phenomenon and calculate observables. First papers about the role of spin fluctuations
by Doniach and Engelsberg \cite{Doniach} and Izuyama {\em et al.} \cite{Japan} 
were published in sixties. 
Till today, fluctuations have been introduced to many-body and phenomenological models and 
a very popular semiempirical theory proposed by Eliashberg \cite{Eliashberg}. 
  
The goal of this work is to include the spin fluctuations into the density functional theory
for superconductors which, in principle,  anables to calculate all material properties, also
in the superconducting state, from first principles. 
The framework of the SCDFT was set up by Oliveira, Gross and Kohn \cite{OGK} in 1988.
Recently, the SCDFT gap equation has been solved numerically for simple metals
\cite{Martin-OGK,Miguel} and MgB$_{2}$ \cite{Massidda}.

As for the critical temperatures, it is known for a long time, that spin fluctuations 
decrease considerably T$_{c}$ of some superconductors  \cite{Berk,Winter}.
In our previous work for niobium under pressure \cite{our-Nb}, we solved the
gap equation of the Eliashberg theory \cite{Eliashberg} with and without spin fluctuations and
the SCDFT gap equation only with the Coulomb and phonon interactions. 
We found that the effect of paramagnons decreased T$_{c}$ obtained from the Eliashberg theory 
by 3-4~K, however, an approximate treatment of the Coulomb interactions by a simple constant, 
$\mu^{\ast}$, led to a large disagreement of the theoretical results with the
experimetal data  \cite{Struzhkin}.
In contrast to the Eliashberg theory, the SCDFT scheme is parameter free, 
but the critical temperature calculated without spin fluctuations 
for Nb at ambient pressure \cite{our-Nb} was about 3.7~K higher than the experimental T$_{c}$. 

In this work, we follow the derivations of the SCDFT gap equation given in a number of
PhD theses\footnote{available at URL: www.physik.fu-berlin.de/$\sim$ag-gross} 
\cite{SK,ML,MM}, and we include the spin fluctuations. The paramagnon spectral function 
is calculated within the random phase approximation (RPA) with the assumption of  
the homogeneous electron gas, similarly to the work by Berk and Schrieffer \cite{Berk} 
done for the Eliashberg theory. 
We solve the obtained gap equation for a few  simple metals and update our previous results 
for niobium under pressure.

In the following Sections, we introduce the SCDFT gap equation and 
the construction of the exchange-correlation functional, $F_{xc}$, 
by collecting the most important building blocks of the theory 
given by its authors \cite{OGK} and first developers \cite{SK,ML,MM,SK-dec,KC}.
These Sections are: II. SCDFT gap equation, III. Exchange-correlation functional, 
and IV. Coulomb interaction and phonons in $F_{xc}$. 
Above Sections are written using the notation according to Parks \cite{Parks,Kazumi} 
and Vonsovsky \cite{Vonsovsky}. This notation is in some points, such as Nambu Green's function 
and the selfenergy, different than the notation previously used for the SCDFT \cite{SK,ML,MM}.
We introduce the spin fluctuations in Sections: V. Paramagnons in $F_{xc}$ and 
VI. Gap equation with paramagnons and implementation details.
We report obtained critical temperatures in Section VII, and we summarize in Section VIII.

%% file: gap.tex
\section{SCDFT gap equation}

In this Section, we wish to guide  the reader, step by step, 
to the gap equation which will be solved at the end of this work 
to calculate the critical temperatures.
We start by bringing the fundaments of the SCDFT \cite{OGK} and 
the main approximations, such as the decoupling of band energies and the 
superconducting gap and a linearization of the gap equation close to $T_{c}$, 
which were assumed for numerical convenience \cite{SK,ML,MM,SK-dec}. 
We believe that these approximations do not cause any significant difference
in the calculated critical temperatures.

Turning to details of the SCDFT, in order to obtain the gap equation 
one needs to follow the points below:

\begin{enumerate}
\item {\em The grand-canonical Hamiltonian} for a superconductor reads
\begin{eqnarray}
&& \hat{H}_{v,\Delta}  =  \sum_{\sigma} \int d^{3}r \; 
\hat{\psi}_{\sigma}^{\dagger}({\bf r})
\left[ -\frac{\nabla^{2}}{2}+v({\bf r})-\mu \right] 
\hat{\psi}_{\sigma}({\bf r}) \nonumber \\ 
&& + \; \frac{1}{2} \int d^{3}r \; d^{3}r' \; 
\hat{\psi}_{\sigma}^{\dagger}({\bf r}) \hat{\psi}_{\sigma}^{\dagger}({\bf r'})
\; \frac{1}{|{\bf r}-{\bf r}'|} \;
\hat{\psi}_{\sigma}({\bf r'})\hat{\psi}_{\sigma}({\bf r})
\nonumber \\
&& -  \; \int d^{3}r_{1} \; d^{3}r'_{1} \; d^{3}r_{2} \; d^{3}r'_{2} \; 
\hat{\psi}_{\downarrow}^{\dagger}({\bf r}'_{1}) 
\hat{\psi}_{\uparrow}^{\dagger}({\bf r}_{1}) \nonumber \\
&& \times \; w({\bf r}'_{1},{\bf r}_{1},{\bf r}_{2},{\bf r}'_{2}) \; 
\hat{\psi}_{\uparrow}({\bf r}_{2})\hat{\psi}_{\downarrow}({\bf r}'_{2}) \nonumber \\
&&  - \left[ \int d^{3}r \; d^{3}r' \; \Delta^{\ast}({\bf r},{\bf r'}) \;
\hat{\psi}_{\uparrow}({\bf r})\hat{\psi}_{\downarrow}({\bf r'})+ H.c. \right],
\end{eqnarray}
where $v({\bf r})$ and $\Delta({\bf r},{\bf r}')$ are  
an external potential and an anomalous pair potential respectively. 
The pairing interaction $w$ in the particular BCS case satisfies 
$w({\bf r}'_{1},{\bf r}_{1},{\bf r}_{2},{\bf r}'_{2}) = 
w({\bf r}'_{1}-{\bf r}_{1},{\bf r}_{2}-{\bf r}'_{2})$.
The normal and anomalous densities, $n({\bf r})$ and $\chi({\bf r},{\bf r'})$, 
are defined as
\begin{eqnarray}
n({\bf r}) & = & \sum_{\sigma} \langle \hat{\psi}_{\sigma}^{\dagger}({\bf r}) 
 \hat{\psi}_{\sigma}({\bf r}) \rangle , \\
\chi({\bf r},{\bf r'}) & = & \langle \hat{\psi}_{\uparrow}({\bf r}) 
 \hat{\psi}_{\downarrow}({\bf r'}) \rangle .
\end{eqnarray}
\item {\em The Hohenberg-Kohn theorem for superconductors} says that, 
 at each temperature $\theta=1/\beta$, the normal and anomalous densities, 
$n({\bf r})$ and $\chi({\bf r},{\bf r'})$, determine uniquely the density operator 
$\hat{\rho}=e^{-\beta \hat{H}_{v,\Delta}} / Tr e^{-\beta \hat{H}_{v,\Delta}} $ 
which minimizes the thermodynamic potential, $\Omega_{v,\Delta} [\hat{\rho}]$, given by 
\begin{equation}
\Omega_{v,\Delta} [\hat{\rho}] = 
Tr \{ \hat{\rho} \; \hat{H}_{v,\Delta} + \theta \; \hat{\rho} \; ln\hat{\rho}  \}.
\end{equation}
\item Furthermore, {\em the thermodynamic potential} 
 can be expressed in terms of the densities and potentials 
by involving a  {\em universal functional} of the densities, 
$F[n,\chi]$, as follows
\begin{eqnarray}
 \Omega_{v,\Delta} [n,\chi]  & = & 
F[n,\chi] + \int d^{3}r \; v({\bf r}) n({\bf r})  \nonumber \\
& - & \int d^{3}r  d^{3}r'  
[ \Delta^{\ast}({\bf r},{\bf r'})\chi({\bf r},{\bf r'}) + H.c. ]. 
\label{omega}
\end{eqnarray}
\item The universal functional contains {\em the exchange-correlation (xc) 
free-energy functional, $F_{xc}[n,\chi]$}, as below 
\begin{eqnarray}
F[n,\chi] & = & T_{s}[n,\chi] - \theta \; S_{s}[n,\chi] -\mu \; N \nonumber \\
& + & \frac{1}{2} \int d^{3}r \; d^{3}r' \; 
\frac{n({\bf r}) n({\bf r'})}{|{\bf r}-{\bf r}'|} \nonumber \\
& - & \int d^{3}r_{1} \; d^{3}r'_{1} \; d^{3}r_{2} \; d^{3}r'_{2} \; 
\chi^{\ast}({\bf r}_{1},{\bf r}'_{1}) \nonumber \\
& \times & w({\bf r}'_{1},{\bf r}_{1},{\bf r}_{2},{\bf r}'_{2})
\chi^{\ast}({\bf r}_{2},{\bf r}'_{2}) + F_{xc}[n,\chi],
\label{free}
\end{eqnarray}
where $T_{s}[n,\chi]$ and $S_{s}[n,\chi]$ are the kinetic energy and the entropy 
of a noninteracting system with the noninteracting  potentials, 
$v_{s}$ and $\Delta_{s}$, 
such that the densities $n$ and $\chi$ are equal to those 
of the noninteracting system. 
In the above formula,  $\mu$ is the chemical potential.
\item {\em The noninteracting grand-canonical Hamiltonian} 
 can be written in terms of the noninteracting densities  and potentials as 
\begin{eqnarray}
 \hat{H}_{s}  & = & \sum_{\sigma} \int d^{3}r \; \hat{\psi}_{\sigma}^{\dagger}({\bf r})
\left[ -\frac{\nabla^{2}}{2}+v_{s}({\bf r})-\mu \right] \hat{\psi}_{\sigma}({\bf r}) 
\nonumber \\
 & - & \left[ \int d^{3}r \; d^{3}r' \; \Delta_{s}^{\ast}({\bf r},{\bf r'})
\hat{\psi}_{\uparrow}({\bf r})\hat{\psi}_{\downarrow}({\bf r'})+H.c. \right].
\end{eqnarray}
\item The diagonalization of the noninteracting Hamiltonian, $\hat{H}_{s}$, 
using the Bogoliubov transformation leads to the 
{\em  Kohn-Sham-Bogoliubov-de Gennes (KS-BdG) equations}
\begin{eqnarray}
\left[ -\frac{\nabla^{2}}{2}+v_{s}({\bf r})-\mu \right] u_{i}({\bf r}) & + &  
                 \int d^{3}r' \; \Delta_{s}({\bf r},{\bf r'}) v_{i}({\bf r'}) \nonumber \\
     & = & E_{i} \; u_{i}({\bf r}), \label{BdG1} \\
-\left[ -\frac{\nabla^{2}}{2}+v_{s}({\bf r})-\mu \right] v_{i}({\bf r}) & + & 
                \int d^{3}r' \; \Delta_{s}^{\ast}({\bf r},{\bf r'}) u_{i}({\bf r'}) \nonumber \\
     & = & E_{i} \; v_{i}({\bf r}),
\label{BdG2}
\end{eqnarray}
with $v_{i}({\bf r})$ and $u_{i}({\bf r})$ being the pair creation and 
anihilation amplitudes respectively.
\item {\em The noninteracting potentials, $v_{s}$ and $\Delta_{s}$,}  
consist of the external potentials, $v_{0}$ and $\Delta_{0}$, 
and Hartree potentials, and the exchange-correlation potentials, 
$v_{xc}$ and $\Delta_{xc}$, as follows 
\begin{eqnarray}
v_{s}[n,\chi]({\bf r}) & = & v_{0}({\bf r}) + 
\int d^{3}r' \; \frac{n({\bf r'})}{|{\bf r}-{\bf r'}|} \nonumber \\
 &  + &  v_{xc}[n,\chi]({\bf r}), \\
\Delta_{s}[n,\chi]({\bf r},{\bf r'}) & = & \Delta_{0}({\bf r},{\bf r'}) + 
\int d^{3}r' \; \frac{\chi({\bf r},{\bf r'})}{|{\bf r}-{\bf r'}|} \nonumber \\
 & + & \Delta_{xc}[n,\chi]({\bf r},{\bf r'}). \label{D0}
\end{eqnarray}
The external pairing potential has been introduced in order to break the symmetry, thus,
in the calculations $\Delta_{0}({\bf r},{\bf r'})\longrightarrow 0$ in Eq. (\ref{D0}).
\item {\em The exchange-correlation potentials, $v_{xc}$ and 
$\Delta_{xc}$,} are  defined as the derivatives of the 
$xc$ functional, $F_{xc}[n,\chi]$,  with respect to the densities, $n$
and $\chi$, correspondingly as below 
\begin{eqnarray}
v_{xc}[n,\chi]({\bf r}) & = & \frac{\delta F_{xc}[n,\chi]}{\delta n({\bf r})}, \\
\Delta_{xc}[n,\chi]({\bf r},{\bf r'}) & = & -\frac{\delta F_{xc}[n,\chi]}
{\delta \chi^{\ast}({\bf r},{\bf r'})}. 
\label{pot-xc}
\end{eqnarray}
\item The densities, $n$ and $\chi$ are defined 
as functions of the amplitudes $u_{i}({\bf r})$ and $v_{i}({\bf r})$ as
\begin{eqnarray}
n({\bf r}) & = & 2 \sum_{i} \; [ \; |u_{i}({\bf r})|^{2}f_{\beta,i} + 
|v_{i}({\bf r})|^{2}(1-f_{\beta,i}) \; ], \\
\chi({\bf r},{\bf r'}) & = &  \sum_{i} \; [ \;
v_{i}^{\ast}({\bf r'})u_{i}({\bf r})(1-f_{\beta,i}) - 
v_{i}^{\ast}({\bf r})u_{i}({\bf r'})f_{\beta,i} \; ], \nonumber \\
&& 
\end{eqnarray}
with the Fermi distribution function $f_{\beta,i}  =  1+ exp(\beta E_{i})$.
\end{enumerate}

At this point, one could guess the densities, $n$ and $\chi$, 
and find the potentials, $v_{xc}$ and 
$\Delta_{xc}$, and solve the KS-BdG equations, 
and find new densities etc. Further for  practical reasons, as we already mentioned 
at the begin of this Section, one can make two approximations which we will discuss now.   
\begin{enumerate}
\addtocounter{enumi}{9}
\item The energy scales for the electronic energies and the superconducting
energy gap differ by orders of magnitude. Therefore, the KS-BdG equations can
be {\em decoupled} into the Kohn-Sham equation and the gap equation. 
This approximation was introduced to the SCDFT in Ref. \cite{SK-dec}. \\

\noindent
It holds within {\em the decoupling approximation} that:
\begin{enumerate}
\item the amplitudes $u_{i}({\bf r})$ and $v_{i}({\bf r})$
can be written in a form 
\begin{eqnarray}
u_{i}({\bf r}) \approx u_{i} \; \varphi_{i}({\bf r}) \; \; ; & \;\; \;\; & 
v_{i}({\bf r}) \approx v_{i} \; \varphi_{i}({\bf r}),
\label{decoupling}
\end{eqnarray}
\item the eigenvalues in Eqs. (\ref{BdG1}) and (\ref{BdG2}) are defined by
\begin{equation}
E_{i}= \pm \sqrt{ \xi_{i}^{2}+| \Delta_{i}}
\end{equation}
where $\xi_{i}=\varepsilon_{i}-\mu$
\item the coefficients $u_{i}$ and $v_{i}$ are given by
\begin{eqnarray}
u_{i} & = & \frac{1}{\sqrt{2}} \; sgn(E_{i}) \;
e^{i\phi_{i}} \; \sqrt{ 1+\frac{\xi_{i}}{E_{i}} }, \\
v_{i} & = & \frac{1}{\sqrt{2}} \; \sqrt{ 1-\frac{\xi_{i}}{E_{i}} }, 
\end{eqnarray}
and the phase factor $\phi_{i}$ is defined by
\begin{equation}
e^{i\phi_{i}} = \frac{\Delta_{i}}{|\Delta_{i}|},
\end{equation}
\item the matrix elements $\Delta_{i}$ are defined as
\begin{equation}
 \Delta_{i} = \int d^{3}r \; \int  d^{3}r' \; \varphi_{i}^{\ast}({\bf r}) 
\Delta_{s} ({\bf r},{\bf r'}) \varphi_{i}({\bf r'}),
\label{deltai}
\end{equation}
\item and the normal and anomalous densities read respectively
\begin{eqnarray}
 n({\bf r})  =  \sum_{i} \left( 1-\frac{\xi_{i}}{E_{i}} \right) 
\tanh\left( \frac{\beta E_{i}}{2} \right) |\varphi_{i}({\bf r})|^{2}, & \;\;\;\;\;\; & 
\label{dens1} \\
 \chi({\bf r},{\bf r'})  =   \frac{1}{2} \sum_{i} \frac{\Delta_{i}}{E_{i}}
\tanh\left( \frac{\beta E_{i}}{2} \right) 
\varphi_{i}({\bf r})\varphi_{i}({\bf r'}). & \;\;\;\;\;\; &
\label{dens2}
\end{eqnarray}
\end{enumerate}
\item The decoupling of the two energy scales yields a transformation of
the KS-BdG equations into {\em the ordinary Kohn-Sham equation} 
\begin{equation}
 - \left[ \frac{\nabla^{2}}{2} + v_{s}[n,\chi] ({\bf r}) - \mu 
\right] \varphi_{i}({\bf r})  =  \epsilon_{i} \; \varphi_{i}({\bf r}), \\
\end{equation}
and {\em the gap equation}
\begin{equation}
  \Delta_{i}  =  \Delta_{Hxc\; i}[\mu,\Delta_{i}]. \label{gap}
\end{equation}
The Eq. (\ref{gap}) stems from including Eqs. (\ref{dens1}) and (\ref{dens2}) 
into Eq. (\ref{D0}), and  using  the potential given by formula (\ref{D0}) 
in Eq. (\ref{deltai}).
\item {\em In vicinity of $T_{c}$}, the gap function is vanishing, therefore, 
it can be {\em linearized} in $\Delta_{i}$.
\end{enumerate}

\noindent
The above twelve steps lead to the gap equation which can be expressed in the form
\begin{eqnarray}
\Delta_{i} & = & -\frac{1}{2} \sum_{j} M_{Hxc,ij}[\mu] \; 
\frac{tanh(\frac{\beta}{2}\xi_{j})}{\xi_{j}}\Delta_{j}, \label{gap-1} \\
M_{Hxc,ij}[\mu] & = & -\frac{\delta \Delta_{Hxc,i}}{\delta\chi_{j}},  
\label{gap-eq}
\end{eqnarray}
where $\Delta_{Hxc,i}$ is defined by Eq. (\ref{pot-xc}). \\
In other way, Eq. (\ref{gap-eq}) can be written as
\begin{equation}
\Delta_{i}  =  - Z_{i}[\mu] \Delta_{i} - \frac{1}{2}
\sum_{j} K_{ij}[\mu] \frac{tanh(\frac{\beta}{2}\xi_{j})}{\xi_{j}}\Delta_{j}.
\label{gap-2}
\end{equation}
$K_{ij}$ and $Z_{i}$ are the functionals only of the chemical potential 
in the case when the gap equation is linearized.
The above gap equation will be solved later in this work. 
The explicit form  of the kernel $K_{ij}$ and the norm $Z_{i}$ 
will be given in Section \ref{implement}. 

Since the gap function (\ref{gap}) contains the exchange-correlation part defined 
by Eq. (\ref{pot-xc}), we will focus on the construction of the exchange-correlation
free-energy functional, $F_{xc}$, in the following Section.

%% file: self.tex
\section{Exchange-correlation functional, $F_{xc}[n,\chi]$}

The derivation of the exchange-correlation energy $F_{xc}$, 
by making use of {\em the perturbative expansion of the thermodynamic potential}, 
was given in Ref. \cite{SK}. 
For the purpose of inclusion the spin interactions, we will briefly draw 
a skeleton of this derivation here.
 
First, one can notice from Eqs. (\ref{omega}) and (\ref{free}) that 
\begin{eqnarray}
F_{xc} & = & \Omega - \Omega_{s} + \int d^{3}r \; [v_{H}({\bf r}) + 
v_{xc}({\bf r})] n({\bf r}) \nonumber \\
& -  & \int d^{3}r \; d^{3}r' \;
[\Delta_{xc}^{\ast}({\bf r},{\bf r'}) \chi({\bf r},{\bf r'}) +
\Delta_{xc}({\bf r},{\bf r'}) \chi^{\ast}({\bf r},{\bf r'})] \nonumber \\
& -  & \frac{1}{2} \int  d^{3}r \; d^{3}r' \;
       \frac{n({\bf r})n({\bf r'})}{|{\bf r}-{\bf r'}|}. 
\label{Fxc}
\end{eqnarray}
 
Then, we take {\em the coupling constant integration formula} which reads 
\begin{equation}
\Omega- \Omega_{s}   =  
 \int_{0}^{1} \frac{d\lambda}{\lambda} 
\langle \lambda \hat{H_{1}} \rangle ,
\label{lambda}
\end{equation}
where $\lambda$ is the coupling constant, and the perturbation Hamiltonian $\hat{H_{1}}$ 
satisfies $\hat{H} = \hat{H_{s}} + \lambda \hat{H_{1}}$
with the interacting and noninteracting Hamiltonians, $\hat{H}$ and $\hat{H_{s}}$, respectively.
The Hamiltonian $\hat{H_{1}}$ contains the difference between the exact Coulomb interaction and 
the exchange-correlation potentials, the electron-phonon interaction, 
the electron-paramagnon interaction etc. \\ 
The average in Eq. (\ref{lambda}) has to be taken with the density operator
$\hat{\rho}_{\lambda}=e^{-\beta H_{\lambda}}/Z_{\lambda}$.

Before an explicit evaluation of the coupling constant integration formula 
(\ref{lambda}), we write here a definition of the Nambu Green's function  
\begin{eqnarray}
&& \bar{G}_{\sigma\sigma'}({\bf r}\tau,{\bf r'}\tau')  =  \nonumber \\
&& \left(
\begin{array}{cc}
G_{\sigma\sigma'}({\bf r}\tau,{\bf r'}\tau') & 
F_{\sigma-\sigma'}({\bf r}\tau,{\bf r'}\tau') \\
F^{\dagger}_{-\sigma\sigma'}({\bf r}\tau,{\bf r'}\tau') & 
-G_{-\sigma'-\sigma}({\bf r'}\tau',{\bf r}\tau) 
\label{Nambu}
\end{array}
\right),
 \end{eqnarray}
which is a 2$\times$2-matrix of the normal and anomalous 
{\em single particle Green's functions}, $G_{\sigma\sigma'}$ and $F_{\sigma\sigma'}$, 
given respectively by
\begin{eqnarray}
 G_{\sigma\sigma'}({\bf r}\tau,{\bf r'}\tau') & = &
- \langle \hat{T} \hat{\psi}_{\sigma}({\bf r}\tau)  
\hat{\psi}^{\dagger}_{\sigma'}({\bf r'}\tau') \rangle , \label{G1} \\
 F_{\sigma\sigma'}({\bf r}\tau,{\bf r'}\tau') & = & 
- \langle \hat{T} \hat{\psi}_{\sigma}({\bf r}\tau)  
\hat{\psi}_{\sigma'}({\bf r'}\tau') \rangle , \label{F1} \\
 F^{\dagger}_{\sigma\sigma'}({\bf r}\tau,{\bf r'}\tau') & = & 
- \langle \hat{T} \hat{\psi}^{\dagger}_{\sigma}({\bf r}\tau)  
\hat{\psi}^{\dagger}_{\sigma'}({\bf r'}\tau') \rangle \label{F2} .
\label{NambuG}
\end{eqnarray} 

The detailed derivation of $\langle \hat{H}_{1}\rangle $ is given in Refs. \cite{SK,ML,MM}. 
This derivation starts from {\em the equations of motion} for the field operator, $\hat{\psi}_{\sigma}$, 
and for the noninteracting Green's function, 
$\bar{G}^{s}_{\sigma\sigma'}$, which are as follows  
\begin{eqnarray}
 \frac{\partial}{\partial \tau} \; \hat{\psi}_{\sigma}({\bf r}\tau) & = & 
e^{\hat{H}\tau} \; [\hat{H},\hat{\psi}_{\sigma}({\bf r})] \; e^{- \hat{H}\tau}, \\
\hat{\cal L} \; \bar{G}^{s}_{\sigma\sigma'}({\bf r}\tau,{\bf r'}\tau') & =&
- \delta_{\sigma\sigma'} \; \delta({\bf r}-{\bf r'}) \; \delta(\tau-\tau'),
\end{eqnarray}
with the Kohn-Sham Hamiltonian for the normal state, $\hat{h}_{s}$, 
and the operator $\hat{\cal L}$ given respectively by 
\begin{eqnarray}
 \hat{h}_{s}({\bf r}) & = & -\frac{\nabla^{2}}{2} + v_{s}({\bf r}) - \mu, \\
 \hat{\cal L} & = & \left(   
\begin{array}{cc}
\frac{\partial}{\partial \tau} + \hat{h}_{s}({\bf r}) &   \hat{\Delta}_{s}({\bf r})  \\
 \hat{\Delta}^{\ast}_{s}({\bf r})   & \frac{\partial}{\partial \tau} - \hat{h}_{s}({\bf r})
\end{array}
\right).
\end{eqnarray}
The  operator $\hat{\Delta}_{s}({\bf r})$ is defined as
\begin{equation}
\hat{\Delta}_{s}({\bf r}) \; f({\bf r}) = 
\int d^{3}r' \; \hat{\Delta}_{s}({\bf r},{\bf r'}) f({\bf r'}).
\end{equation}
In order to complete the derivation, 
one also needs to make use of {\em the Dyson's equation}
\begin{eqnarray}
 \bar{G}_{\sigma'\sigma}({\bf r}\tau,{\bf r'}\tau') & = &
\bar{G}^{s}_{\sigma'\sigma}({\bf r}\tau,{\bf r'}\tau') \nonumber \\
& + &  \sum_{\sigma\sigma'} \int d^{3}r_{1} \; d^{3}r_{2} 
\int d\tau_{1} \; d\tau_{2} \; 
\bar{G}^{s}_{\sigma\sigma_{1}}({\bf r}\tau,{\bf r}_{1}\tau_{1}) \nonumber \\
& \times & \bar{\Sigma}({\bf r}_{1}\tau_{1},{\bf r}_{2}\tau_{2}) \;
\bar{G}_{\sigma_{2}\sigma'}({\bf r}_{2}\tau_{2},{\bf r'}\tau'),
\end{eqnarray}
with $\bar{\Sigma}$ being the self-energy. \\ 

The above building blocks make us to arrive, after some algebra, at the relation 
\begin{eqnarray}
\Omega- \Omega_{s}  & = & \frac{1}{2} \int_{0}^{1} \frac{d\lambda}{\lambda} 
\left\{ \sum_{\sigma\sigma'} \int d^{3}r \; d^{3}r' \right. \nonumber \\
 &  \times & \int d\tau' \; [ \bar{\Sigma}_{\sigma\sigma'}^{\lambda}({\bf r}\tau,{\bf r'}\tau')  
\bar{G}_{\sigma'\sigma}^{\lambda}({\bf r'}\tau',{\bf r}\tau^{+})]_{11}  \nonumber \\
 & - & \lambda \int d^{3}r \; [v_{H}({\bf r}) + 
              v_{xc}({\bf r})] \; n^{\lambda}({\bf r}) \nonumber \\
 & +  & \left.  2\lambda \int d^{3}r \; d^{3}r' \;
    \Delta_{xc}^{\ast}({\bf r},{\bf r'}) \chi^{\lambda}({\bf r},{\bf r'}) \right\}, 
\label{grancan}
\end{eqnarray}
which we can plug into the Eq. (\ref{Fxc}) for the exchange-correlation functional, $F_{xc}[n,\chi]$. \\ 

As for {\em the first-order selfenergy}, $\bar{\Sigma}_{\sigma\sigma'}$, for the nonmagnetic
systems with the potential $v({\bf r}\tau,{\bf r'}\tau')$, this energy is defined as
\begin{equation}
\bar{\Sigma}({\bf r}\tau,{\bf r'}\tau')   =  
- v({\bf r}\tau,{\bf r'}\tau') \; 
{\tau}_{3}\bar{G}({\bf r}\tau,{\bf r'}\tau') 
{\tau}_{3}, \label{self} 
\end{equation}
and $\tau_{3}$ is one of the Pauli matrices:
\begin{eqnarray}
\tau_{1} = \left( 
\begin{array}{cc}
0  & 1 \\
1 &  0
\end{array}
\right), &
\tau_{2} = \left( 
\begin{array}{cc}
0  & i \\
-i &  0
\end{array}
\right), \nonumber \\
\tau_{3} = \left( 
\begin{array}{cc}
1  & 0 \\
0 &  -1
\end{array}
\right), &
\tau_{0} = \left( 
\begin{array}{cc}
1  & 0 \\
0 &  1
\end{array}
\right). \nonumber
\end{eqnarray}
For the magnetic systems, the matrix $\tau_{3}$ in the each vertex of Feynman diagrams for the selfenergy
with  the Coulomb and phonon interactions has to be replaced with the matrix $\tau_{0}\tau_{3}$. \\

In this Section, we sketched main steps to be done for finding a general form of 
the  $F_{xc}[n,\chi]$ functional for a superconductor. 
The final formula involves the selfenergy which will be evaluated
in detail for the Coulomb and electron-phonon interactions in the next Section and for
the paramagnons in Section \ref{para}.

%% file: normal.tex
\section{Coulomb and electron-phonon interactions in $F_{xc}[n,\chi]$}

The derivation of $F_{xc}$ for the Coulomb and phonon interactions 
is given in detail in Refs. \cite{Martin-OGK,ML}. 
Here, we report this derivation starting with the interactions in the selfenergy 
(in Eq. (\ref{self})) defined by
\begin{eqnarray}
v^{el}({\bf r},{\bf r'}) & = & \frac{1}{|{\bf r}-{\bf r}'|}, \\
v^{ph}({\bf r}\tau,{\bf r'}\tau') & = &  
V_{\lambda{\bf q}}({\bf r}) D_{\lambda{\bf q}}(\tau-\tau') 
V_{\lambda{\bf q}}({\bf r'}),
\end{eqnarray}
where $V_{\lambda{\bf q}}$ is the electron-phonon interaction vertex and 
$D_{\lambda{\bf q}}$ is the phonon Green's function defined as
\begin{equation}
D_{\lambda{\bf q}} (\tau,\tau') = 
\langle \hat{T} \hat{\Phi}_{\lambda{\bf q}}(\tau) 
\hat{\Phi}^{\dagger}_{\lambda{\bf q}}(\tau') \rangle,
\end{equation} 
with $\hat{\Phi}_{\lambda{\bf q}}=b_{\lambda,{\bf q}}+b^{\dagger}_{\lambda,{\bf -q}}$, 
and $b^{\dagger}_{\lambda,{\bf q}}$ ($b_{\lambda,{\bf q}}$) being
the phonon creation (anihilation) operators. \\

Let us have a look now at the expression (\ref{Fxc}) for $F_{xc}$ and the definitions
of the Nambu Green's function and selfenergy given by Eqs. (\ref{Nambu}) and 
(\ref{self}) respectively.
The (1,1)-element of the ($\bar{\Sigma}\bar{G}$)-matrix, present in 
the formula (\ref{grancan}) and enterring Eq. (\ref{Fxc}),
is proportional to 
\begin{equation}
G_{\uparrow\uparrow}G_{\uparrow\uparrow} - 
F_{\uparrow\downarrow}F^{\dagger}_{\uparrow\downarrow} =
G_{\uparrow\uparrow}G_{\uparrow\uparrow} + 
F_{\uparrow\downarrow}F^{\dagger}_{\downarrow\uparrow}
\label{signum}
\end{equation}
and the corresponding terms with the opposite spins. 
The above terms appear for both the Coulomb and electron-phonon interactions, and
later will lead to the opposite signums in the kernel $K_{ij}$ and the norm $Z_{i}$ 
of the gap equation. Just mentioned difference in signum, in the first order terms of the total energy 
with the normal and anomalous Green's functions, stems from the factor of (-1) which one 
has to associate with the each loop of anomalous Green's functions. \\
 
In order to evaluate further $F_{xc}$, we bring here the explicite expressions for
the noninteracting propagators.
The formulas given below were derived from the definitions (\ref{G1}-\ref{F2}) 
assumming the decoupling approximation, {\em i.e.} Eqs. (\ref{decoupling});
the Kohn-Sham orbitals $\varphi_{\bf k}({\bf r})$ were chosen to those of 
a homogeneous gas ($w_{n}$ are the odd Matsubara frequencies) 
\begin{eqnarray}
&& G^{s}_{\sigma\sigma'}({\bf k},w_{n})  = \delta_{\sigma,\sigma'} \nonumber \\
&& \;\;\;\;\;\;\;\;\;\;\;\;\; 
\times \; \left[ \frac{|u_{{\bf k}}|^{2}}{i\omega_{n}-E_{{\bf k}}} +
\frac{|v_{{\bf k}}|^{2}}{i\omega_{n}+E_{{\bf k}}}  \right], \label{G1d} \\
&& F^{s}_{\sigma\sigma'}({\bf k};w_{n})  = \delta_{\sigma,-\sigma'} \; sgn(\sigma') \; \nonumber \\
&& \;\;\;\;\;\;\;\;\;\;\;\;\; 
\times \; u_{{\bf k}}v^{\ast}_{{\bf k}} \; \left( 
\frac{1}{i\omega_{n}+E_{{\bf k}}} - \frac{1}{i\omega_{n}-E_{{\bf k}}}  \right),  \label{F1d} \\
&& F^{s\dagger}_{\sigma\sigma'}({\bf k};w_{n})  = \delta_{\sigma,-\sigma'} \;  sgn(\sigma) 
\; \nonumber \\
&& \;\;\;\;\;\;\;\;\;\;\;\;\; 
\times \; u^{\ast}_{{\bf k}}v_{{\bf k}} \; \left( 
\frac{1}{i\omega_{n}+E_{{\bf k}}} - \frac{1}{i\omega_{n}-E_{{\bf k}}}  \right). \label{F2d} 
\end{eqnarray}   \\

Now, we will combine Eqs. (\ref{Fxc}) and (\ref{grancan}), for the $F_{xc}$ and $\Omega-\Omega_{s}$ respectively, 
with a definition of the Nambu Green's function, Eq. (\ref{Nambu}), and an expression for 
the selfenergy, Eq. (\ref{self}). As for the noniteracting Green's functions, we use those obtained 
within the decoupling approximation, {\em i.e.} (\ref{G1d}-\ref{F2d}). 
This way, one arrives to formulas for the xc energy, stemming from the normal and anomalous loops,
which we write here. 
The "normal" and "anomalous" terms of $F_{xc}$ for the  electronic contributions,  
$F_{xc}^{el,1}$ and $F_{xc}^{el,2}$, are as follows 
\begin{eqnarray}
F_{xc}^{el,1} & = &  -\frac{1}{4} \sum_{\bf kk'} \left( 1-\frac{\xi_{\bf k}}{E_{\bf k}}\right) 
v({\bf k},{\bf k'}) \left( 1-\frac{\xi_{\bf k'}}{E_{\bf k'}}\right) \nonumber \\
&& \times \tanh\left( \frac{\beta}{2} E_{\bf k}\right) 
\tanh\left( \frac{\beta}{2} E_{\bf k'}\right), \label{el1} \\
F_{xc}^{el,2} & = &  \frac{1}{4} \sum_{\bf kk'}  v({\bf k},{\bf k'})
\frac{\Delta_{\bf k}}{E_{\bf k}}\frac{\Delta_{\bf k'}}{E_{\bf k'}} \nonumber  \\
&& \times \tanh\left( \frac{\beta}{2} E_{\bf k}\right) \tanh\left( \frac{\beta}{2} E_{\bf k'}\right), 
\label{el2} 
\end{eqnarray}
and the electron-phonon terms, 
with the normal and anomalous loops, $F_{xc}^{ph,1}$ and $F_{xc}^{ph,2}$i, respectively
are given below
\begin{eqnarray}
F_{xc}^{ph,1} & = & 
- \frac{1}{2} \; \sum_{\bf kk'} \; \int d\Omega \; \alpha^{2}F(\Omega) \; \nonumber \\
&& \times \; \left[  \left( 1+\frac{\xi_{\bf k}\xi_{\bf k'}}{E_{\bf k}E_{\bf k'} }  \right)
I(E_{\bf k},E_{\bf k'},\Omega) \right. \nonumber \\
&& + \left. \left( 1-\frac{\xi_{\bf k}\xi_{\bf k'}}{E_{\bf k}E_{\bf k'} } \right)
I(E_{\bf k},-E_{\bf k'},\Omega) \right], \label{ph1} \\ 
F_{xc}^{ph,2} & = & \frac{1}{2} \; \sum_{\bf kk'} \; \int d\Omega \; \alpha^{2}F(\Omega) \; 
 \frac{\Delta_{\bf k}\Delta^{\ast}_{\bf k'}}{E_{\bf k}E_{\bf k'}} \nonumber \\  
&& \times \; [I(E_{\bf k},E_{\bf k'},\Omega)-I(E_{\bf k},-E_{\bf k'},\Omega)]. \label{ph2}  
\end{eqnarray}
The function $I(E_{\bf k},E_{\bf k'},\Omega)$ is defined as
\begin{eqnarray}
I(E_{\bf k},E_{\bf k'},\Omega) & = & \frac{1}{\beta^{2}} \sum_{\omega_{1}\omega_{2}}
\frac{1}{i\omega_{1}-E_{\bf k}} \frac{1}{i\omega_{2}-E_{\bf k'}} \nonumber \\
&& \times \; \frac{-2\Omega}{(\omega_{1}-\omega_{2})^{2}+\Omega^{2}} . 
\label{Ifunc}
\end{eqnarray}

For the completeness, we give the definitions:
\begin{eqnarray}
v({\bf k},{\bf k'}) & = & \int d^{3}r \; d^{3}r' \; 
\varphi^{\ast}_{\bf k}({\bf r}) \varphi_{\bf k}({\bf r'})  \nonumber \\
&& \times \; \frac{1}{|{\bf r}-{\bf r'}|} \;
\varphi^{\ast}_{\bf k'}({\bf r}) \varphi_{\bf k'}({\bf r'}), \\
g^{\lambda {\bf q}}_{\bf k,k+q} & = &  \int d^{3}r \; 
\varphi^{\ast}_{\bf k}({\bf r}) \;
  V_{\lambda {\bf q}} \; \varphi_{\bf k+q}({\bf r}), \\
\alpha^{2}F(\Omega) & = & \frac{1}{N(\varepsilon_{F})} 
\sum_{\lambda {\bf q}} \sum_{\bf k} \;
|g^{\lambda {\bf q}}_{\bf k,k+q}|^{2} \;
\delta(\Omega-\omega_{\lambda{\bf q}}) \nonumber \\
&& \times \; \delta(\varepsilon_{\bf k}-\varepsilon_{F})
\; \delta(\varepsilon_{\bf k+q}-\varepsilon_{F}),
\end{eqnarray}
where $\omega_{\lambda{\bf q}}$ is the phonon frequency and 
$N(\varepsilon_{F})$ is the density of states. \\

Using the formulas (\ref{el1}-\ref{Ifunc}), one is ready to derive 
the exchange-correlation potential defined by Eq. (\ref{pot-xc}). 
This derivation can be performed with the help of {\em the chain rule}
as follows 
\begin{eqnarray}
\Delta_{xc,i} & = & - \frac{\delta F_{xc}}{\delta \mu} 
\frac{\delta \mu}{\delta \chi_{i}^{\ast}} - \sum_{j} \left[ 
\frac{\delta F_{xc}}{\delta |\Delta_{j}|^{2}} 
\frac{\delta |\Delta_{j}|^{2}}{\delta \chi_{i}^{\ast}}
  \right. \nonumber \\
& & + \left. \frac{\delta F_{xc}}{\delta (\phi_{j})} 
\frac{\delta (\phi_{j})}{\delta \chi_{i}^{\ast}} \right].
\label{chain}   
\end{eqnarray}
Further evaluation of the above expression is given in detail 
in Refs. \cite{Martin-OGK,SK}. In this work, we give  
the final formula for $\Delta_{xc,i}$ which involves 
the phonon and paramagnon spectral functions 
and can be implemented in a straightforward way.
We will give the details of implementation in Section \ref{implement}. \\

At this point, we arrived to the explicit expressions for $F_{xc}$ with the
electronic and phononic parameters such as: 
the chemical potential $\mu$, the density of states $N(\varepsilon_{F})$,
the single particle energies $\varepsilon_{\bf k}$, 
and the Eliashberg function $\alpha^{2}F(\Omega)$.
Now, we are ready to introduce the spin fluctuations into the discussed formalism,
and we will this in the following Section.

%% file: magnetic.tex
\section{Paramagnons in $F_{xc}[n,\chi]$}
\label{para}

We will introduce the transverse spin-fluctuations to the total energy 
within the SCDFT. For the simplicity, we will assume the  singlet pairing and 
the $s$-wave symmetry of the gap function.
The extension to triplet superconductors could be done following the work by 
Capelle {\em et al.} \cite{KC,triplet}. 
For the case of magnetic superconductors, one should take also into account 
a correction for the Zeeman effect, {\em i.e.} the spin gap. 
As for the paring potentials with the higher angular-momentum, 
one cannot average spherically the angular part 
of the interaction in the RPA formula for the paramagnon susceptibility, 
which formula will be used later in this Section.   \\

Here, we start with the Nambu Green's function for 
the superconductors with magnetic interactions included into the description.
This matrix is now 4$\times$4 dimentional and reads 
\begin{equation}
 \bar{G}({\bf r}\tau,{\bf r'}\tau') = -\langle 
\hat{T}\hat{\Psi}^{\dagger}({\bf r},\tau) \otimes 
\hat{\Psi}({\bf r'},\tau') \rangle ,
\label{Nambu-par}
\end{equation}
with the 4-component field operators 
(the notation has been chosen according to Maki in Ref. \cite{Kazumi} and $x$ denotes 
the vector ({\bf r},$\tau$))
\begin{eqnarray}
\hat{\Psi}(x) = \left( 
\begin{array}{c}
\hat{\psi}_{\uparrow}(x) \\
\hat{\psi}_{\downarrow}(x) \\
\hat{\psi}^{\dagger}_{\uparrow}(x) \\
\hat{\psi}^{\dagger}_{\downarrow}(x)
\end{array} \right), & &
\hat{\Psi}^{\dagger}(x) = \left( 
\hat{\psi}^{\dagger}_{\uparrow}(x) \hat{\psi}^{\dagger}_{\downarrow}(x) 
\hat{\psi}_{\uparrow}(x) \hat{\psi}_{\downarrow}(x) \right). \nonumber \\
\end{eqnarray}

The first-order selfenergy with the spin dependent interaction $v^{\mu\nu}$,
where $\mu$ and $\nu$ denote the cartesian components of the spin orientations of
two interacting electrons, is given by
\begin{eqnarray}
 \bar{\Sigma}({\bf r}\tau,{\bf r'}\tau') & = & 
- v^{\mu\nu}({\bf r}\tau,{\bf r'}\tau') \; 
\hat{\alpha}_{\mu}\bar{G}({\bf r}\tau,{\bf r'}\tau') 
 \hat{\alpha}_{\nu}, \;\;\;  \label{magn-self} \\
v^{\mu\nu}({\bf r}\tau,{\bf r'}\tau') & = &  
I_{ex}({\bf r}) D^{\mu\nu}(\tau-\tau') 
I_{ex}({\bf r'}).  \;\;\;\ \label{pot-magn}
\end{eqnarray}
The quantity $I_{ex}$ is {\em the spin exchange interaction}, 
and $D^{\mu\nu}$ is {\em the spin Green's function}. 
The matrix $\hat{\alpha}_{\mu}$ is defined as
\begin{equation}
\hat{\alpha}_{\mu} = \left( 
\begin{array}{cc}
\sigma_{\mu} & 0 \\
0 & -\sigma_{\mu}^{tr} 
\end{array}
\right),
\label{alfa}
\end{equation}
where $\sigma_{\mu}^{tr}$ denotes a matrix transposed to
the Pauli matrix $\sigma_{\mu}$ (see Ref.~\cite{Vonsovsky}). \\
 
For the transverse spin fluctuations, the $\alpha$-matrix, given by formula (\ref{alfa}), 
involves the Pauli matrices $\sigma^{+}$ and $\sigma^{-}$ defined as 
$\sigma^{\pm}=\frac{1}{2}(\sigma_{x}\pm i\sigma_{y})$; explicitely
\begin{eqnarray}
\sigma^{+} = \left( 
\begin{array}{cc}
0 & 1 \\
0 & 0
\end{array}
\right), & \; &
\sigma^{-} = \left( 
\begin{array}{cc}
0 & 0 \\
1 & 0
\end{array}
\right). \nonumber
\end{eqnarray} \\

Evaluation of the selfenergy with paramagnons, 
according to Eqs. (\ref{magn-self}-\ref{alfa}), yields 
a very sparse 4$\times$4-matrix which reads 
\begin{eqnarray}
& & \bar{\Sigma}({\bf r}\tau,{\bf r'}\tau') 
 =  -v^{+-}({\bf r}\tau,{\bf r'}\tau') \nonumber \\
&&  \times \left( 
\begin{array}{cccc}
G_{\downarrow\downarrow}({\bf r}\tau,{\bf r'}\tau') & 0 & 0 
& -F_{\downarrow\uparrow}({\bf r}\tau,{\bf r'}\tau') \\
0 & 0 & 0 & 0 \\
0 & 0 & 0 & 0 \\
-F^{\dagger}_{\uparrow\downarrow}({\bf r}\tau,{\bf r'}\tau') & 0 & 0 
& G^{\dagger}_{\uparrow\uparrow}({\bf r}\tau,{\bf r'}\tau')
\end{array}
\right),  \nonumber \\
 && \label{self-matrix}
\end{eqnarray}
where $G^{\dagger}_{\uparrow\uparrow}=-G_{\uparrow\uparrow}$. \\

Now, if we go back to the previous section and look again at the 
(1,1)-element of the ($\bar{\Sigma}\bar{G}$)-matrix, we will remind to us that 
for the Coulomb and electron-phonon interactions, the total energy is
proportional to the expression (\ref{signum}).
For the magnetic interactions, however, for which the Nambu Green's function 
has been defined by Eq. (\ref{Nambu-par}) and the selfenergy has been given by 
Eq. (\ref{self-matrix}), the total energy is proportional to
\begin{equation}
G_{\downarrow\downarrow}G_{\uparrow\uparrow} - 
F_{\downarrow\uparrow}F^{\dagger}_{\uparrow\downarrow} =
G_{\downarrow\downarrow}G_{\uparrow\uparrow} - 
F_{\uparrow\downarrow}F^{\dagger}_{\downarrow\uparrow}. 
\label{sfetot}
\end{equation}
The above expression differs from relation (\ref{signum}) by signum in front of
the anomalous Green's functions. 
This difference will show up in the kernel $K_{ij}$ and the norm $Z_{i}$ of the gap equation
such that, both the phonon and paramagnon spectral functions enter the kernel with different
signum (originating from the anomalous loop of Green's functions) 
and the norm with the same signum (originating from the normal loop). \\

To proceed further with the evaluation of the {\em xc}-free energy, $F_{xc}$, 
we write explicitely  the spin-fluctuation Green's function, $D^{\mu\nu}(\tau-\tau')$,
used in Eq. (\ref{pot-magn}). In the case of paramagnons, $D^{\mu\nu}(\tau-\tau')$ is 
the transverse spin susceptibility, $\chi^{+-}$, defined as
\begin{equation}
\chi^{+-}({\bf r}-{\bf r}',\tau-\tau') = \langle \hat{T} \hat{S}^{-}({\bf r},\tau)
\hat{S}^{+}({\bf r'},\tau') \rangle, 
\end{equation}
with the operators increasing and lowerring spin which are defined respectively as
\begin{eqnarray}
\hat{S}^{+}({\bf r},\tau) & = & \hat{\psi}^{\dagger}_{\uparrow}({\bf r},\tau) 
\hat{\psi}_{\downarrow}({\bf r},\tau), \\
\hat{S}^{-}({\bf r},\tau) & = & \hat{\psi}^{\dagger}_{\downarrow}({\bf r},\tau) 
\hat{\psi}_{\uparrow}({\bf r},\tau). 
\end{eqnarray}

For the conduction band, we can use a model of the homogeneous electron gas 
with the fluctuations treated on the level of the random phase approximation. 
The Fourier transform of the RPA-"dressed" paramagnon propagator is
\begin{equation}
\chi^{+-}({\bf q},\nu_{n}) = 
\frac{\chi^{0}({\bf q},\nu_{n})}{1-I_{ex}\chi^{0}({\bf q},\nu_{n})},
\end{equation}
with the Pauli susceptibility $\chi^{0}$ and the even Matsubara frequencies $\nu_{n}$.

It is convenient to introduce the spectral representation   
\begin{eqnarray}
\chi^{+-}({\bf q},\nu_{n}) & = & 
-\int_{0}^{\infty} \frac{d\Omega}{\pi} \; D^{0}(\Omega,\nu_{n}) \; 
\Im m \; \chi^{+-}({\bf q},\Omega), \nonumber \\
  & & \\
D^{0}(\Omega,\nu_{n}) & = & \frac{-2\Omega}{\nu_{n}^{2}+\Omega^{2}},
\end{eqnarray}
and the momentum averaged paramagnon spectral function
\begin{eqnarray}
P(\Omega) & = & N(\varepsilon_{F}) \; \int_{0}^{2k_{F}} dq \; 
\frac{q}{2k_{F}^{\;\; 2}} \; \nonumber \\
& & \times \; |I(q)|^{2} \; \left[ - \frac{1}{\pi} \; 
\Im m \; \chi^{+-} (q,\Omega) \right].  
\end{eqnarray}     
We assume that the interaction function, $I(q)$, is the momentum independent quantity 
$I_{ex}$, which can be calculated in a way given for instance in Ref. \cite{our-Nb}.

Therefore, for the systems with the electron-paramagnon interactions, 
the exchange-correlation free energy  is given by
\begin{eqnarray}
F_{xc}^{sf,1} & = & - \frac{1}{2} \; \sum_{\bf kk'} \; \int d\Omega \; P(\Omega) \; \nonumber \\
&& \times \; \left[  \left( 1+\frac{\xi_{\bf k}\xi_{\bf k'}}{E_{\bf k}E_{\bf k'} }  \right)
I(E_{\bf k},E_{\bf k'},\Omega) \right. \nonumber \\
&& + \left. \left( 1-\frac{\xi_{\bf k}\xi_{\bf k'}}{E_{\bf k}E_{\bf k'} } \right)
I(E_{\bf k},-E_{\bf k'},\Omega) \right], \label{sf1} \\
F_{xc}^{sf,2} & = & -\frac{1}{2} \; \sum_{\bf kk'} \; \int d\Omega \; P(\Omega) \; 
 \frac{\Delta_{\bf k}\Delta^{\ast}_{\bf k'}}{E_{\bf k}E_{\bf k'}} \nonumber \\  
&& \times \; [I(E_{\bf k},E_{\bf k'},\Omega)-I(E_{\bf k},-E_{\bf k'},\Omega)], \label{sf2} 
\end{eqnarray}
where the function $I(E_{\bf k},E_{\bf k'},\Omega)$ is defined by Eq.~(\ref{Ifunc}). 
The explicit formula for the paramagnon spectral function, $P(\Omega)$, within the RPA is
given for instance in Refs. \cite{Berk,Vonsovsky,our-Nb}.

%% file: implementation.tex
\section{Gap equation with paramagnons and implementation details}
\label{implement}

At this point, when we have completed the derivation of all components of the exchange-correlation free
energy: the Coulomb part \-- Eqs. (\ref{el1}-\ref{el2}), the phonon part \-- Eqs. (\ref{ph1}-\ref{ph2}),
and the spin-fluctuation part \-- Eqs. (\ref{sf1}-\ref{sf2}), we can 
write explicitely the gap equation given by  Eqs. (\ref{gap-1}-\ref{gap-2}).

The $M_{ij}$-matrix of the linearized equation (\ref{gap-1}) is the following function of 
the kernel $K_{ij}$ and the norm $Z_{i}$ 
\begin{equation}
 M_{ij}  = 
-\frac{1}{2} \frac{K_{ij}\left[ \Delta=0 \right]}{1-Z_{i}\left[ \Delta=0 \right]}. 
              \label{kernel}  
\end{equation}
The nondiagonal part of the $M_{ij}$-matrix is given by
\begin{equation}
 K_{ij}  = K_{ij}^{el} + K_{ij}^{ph+sf}, 
\end{equation}
where the electronic part is defined by 
\begin{eqnarray}
 K_{ij}^{el} & = & w_{ij}, \\
w_{ij} & = & \frac{2\pi}{k_ik_j} \; log \left(  
    \frac{(k_i+k_j)^{2}+k_{TF}^{2}}{(k_i-k_j)^{2}+k_{TF}^{2}} \right). 
\end{eqnarray}
The Coulomb interaction $w_{ij}$ has been spherically averaged over the angular
coordinates since, as we said before, we assumed the $s$-wave pairing.
The electron correlations are taken into account by
the Thomas-Fermi screening constant, $k_{TF}$,
and $k_{i}$ is an absolute value of the reciprocal vector. \\
The electron-phonon and -paramagnon interaction diagonal part of 
the $M_{ij}$-matrix is given by
\begin{eqnarray} 
 K_{ij}^{ph+sf} & = & 
\frac{2}{tanh(\beta\xi_{i} /2)tanh(\beta\xi_{j} /2)} \nonumber \\
   && \times \int d\Omega \; \left[ \alpha^{2}F(\Omega)-P(\Omega) \right] \nonumber \\
   &  &  \times \left[ 
     I(\xi_{i},\xi_{j},\Omega) - I(\xi_{i},-\xi_{j},\Omega) \right].  
\end{eqnarray}
The diagonal part of the $M_{ij}$-matrix is
\begin{equation}
 Z_{i}   =   Z_{i}^{el} + Z_{i}^{ph+sf},    
\end{equation}
where the purely electronic part is
\begin{eqnarray}
 Z_{i}^{el} & = & -\frac{1}{2\xi_{i}}
   \left\{ \sum_{j} w_{ij} 
   \left[ 1-tanh(\beta\xi_{j}/2) \right]  
    \right. \nonumber \\
   &   &  - \left. \frac{\sum_{jk} 
   \frac{\beta w_{jk}/2} {cosh^{2}(\beta\xi_{j}/2)} 
         \left[ 1-tanh(\beta\xi_{k}/2) \right]} 
        {\sum_{k} \frac{\beta/2}{cosh^{2}(\beta\xi_{k}/2)}}  
                    \right\}, \label{Zel}      
\end{eqnarray}
and the phononic and paramagnon part is
\begin{eqnarray}
Z_{i}^{ph+sf} & = & 
   \frac{-4\pi}{tanh(\beta\xi_{i}/2)}  \frac{1}{\beta} 
  \int d\Omega \; \left[ \alpha^{2}F(\Omega)+P(\Omega) \right]  \nonumber \\ 
  &   &  \times  \sum_{\omega_{2}} \omega_{2} \; sgn(\omega_{2})  
     \left[ Z_{i,sym}^{ph+sf} + Z_{i,asym}^{ph+sf} \right],   \\
 Z_{i,sym}^{ph+sf} & = &  \left[  n_{\beta}(\Omega) + f_{\beta}(-\xi_{i})  \right] 
     \frac{2(\xi_{i}+\Omega)}{\left[ \omega_2^{2}+(\xi_{i}+\Omega)^{2}\right]^{2}}  
                       \nonumber \\
       &   & + \left[  n_{\beta}(\Omega) + f_{\beta}(\xi_{i})  \right] 
   \frac{2(\xi_{i}-\Omega)}{\left[\omega_{2}^{2}+(\xi_{i}-\Omega)^{2}\right]^{2}}.
\end{eqnarray}
Functions $f_{\beta}$ and $n_{\beta}$ are the Fermi-Dirac and 
Bose-Einstein distribution functions respectively.

For the electronic part of the norm, {\em i.e.} $Z_{i}^{el}$,  
we used the zero temperature approximation  
given in Refs. \cite{MM,our-Nb}. This approximation can be justified by the fact that 
the critical temperatures of simple metals, which we calculate in this work, 
are very low. The above simplification is done for sake of a numerical convenience since
there are many singularities in the formula (\ref{Zel}). 

The subscripts "sym" and "asym" mean the symmetric and antisymmetric part of $Z_{i}$ 
with respect to the electron-phonon coupling elements $g_{{\bf k},{\bf k}+{\bf q}}$.
The electron-paramagnon interaction constant, $I({\bf q})$, has been also 
averaged in {\bf q} leading to $I_{ex}$.  
The antisymmetric part $Z_{i,asym}^{ph+sf}$ is ommited in our calculations according 
to the reasons discussed in Refs. \cite{Martin-OGK,MM} and in our previous work \cite{our-Nb}. 
Therefore, we do not give the expression for $Z_{i,asym}^{ph+sf}$ in this work.

%% file: results.tex
\section{Critical temperatures of simple metals}

In the following two subsections, we report the critical
temperatures obtained by solving the SCDFT gap equation with spin fluctuations included.
We compare these results with the results without spin fluctuations and 
results from the Eliashberg theory. First, we calculate parameters of 
the gap equation for several simple metals: V, Mo, Ta, and Pd (fcc and bcc)
at ambient pressure. At the end, we complete our previous results for Nb under pressure
\cite{our-Nb} reporting $T_{c}$ obtained within the SCDFT with the paramagnons included.

The electronic structures, the densities of states (DOS) and the electron-phonon coupling constants 
and the phonon and magnon spectral functions for studied metals were calculated within the local density
approximation (LDA). We used the pseudopotential plane wave codes 
{\sc pwscf} \cite{PWscf} and {\sc espresso} \cite{espresso}. 
The phonons and electron-phonon couplings were obtained from the density functional
perturbation theory (DFPT) \cite{Review}. Since the calculation of the spectral 
function $\alpha^{2}F$ is very time consuming, we used the ultrasoft pseudopotentials
(US PPs) \cite{Vanderbilt}. The kinetic energy cut-offs for 
the wavefunctions and densities were 45~Ry and 270~Ry respectively in order to reproduce
well all features of the phonon dispersions especially for the low frequency phonons 
(see Ref. \cite{our-Nb}).
The metallic broadening at the Fermi energy \cite{Paxton} was assumed at 0.03 Ry.
We used the Monkhorst-Pack mesh \cite{mesh} of (64,64,64)-points for the DOS calculations, 
(16,16,16)-points for the self-consistent calculation of 
the electron-phonon-coupling matrix elements for the each phonon,
the mesh of (8,8,8)-points to fit the phonon dispersions, and the fit from (16,16,16) into 
(64,64,64) mesh-points to perform the integrands with the double-delta function present in
the definition of the electron-phonon coupling constant, 
$\lambda^{ph}$, and the spectral function, $\alpha^{2}F(\omega)$. 

The spin-exchange interaction contants, $I_{ex}$, for metals at ambient
pressure were taken from the work by Singalas {\em et al.} \cite{Singalas}, and further 
we used them for the calculation of the spectral functions, $P(\omega)$, 
and the electron-paramagnon coupling constant, $\lambda^{sf}$. For niobium under pressure,
we used $I_{ex}$ and $P(\omega)$ calculated in our previous work \cite{our-Nb}.       

All electronic parameters and the phonon and magnon spectral functions 
were assumed to be the same for the normal and superconducting state.
The accuracy of functions $\alpha^{2}F(\omega)$ and $P(\omega)$ is very important
for an exact estimation of the critical temperature. The electron-phonon
spectral function, very time consumming for calculations, 
contains all the specific information about the studied system. 
In contrast to $\alpha^{2}F(\omega)$, 
the approximation which we used for the paramagnon spectral function, to avoid calculation of 
this quantity from the time-dependent density functional theory, 
is insuficient. We made the assumption of the homogeneous electron gas for the spin susceptibility and 
the only spin-dependent quantity which we calculated specifically for a given metal was the exchange constant. 
The calculation of this constant, {\em i.e.} $I_{ex}$,
is very difficult and obtained results have a large error due to their very small values and 
necessity to calculate a response function to small magnetizations applied to the system.    
Therefore, as we will see below, the obtained critical temperatures are not always very close 
to the experimental ones. Further development should be directed into more accurate calculation
of the spectral functions, especially $P(\omega)$.   

\subsection{Transition metals at ambient pressure}

\begin{figure} \epsfxsize=8cm \centerline{
\includegraphics[scale=0.30,angle=-90.0]{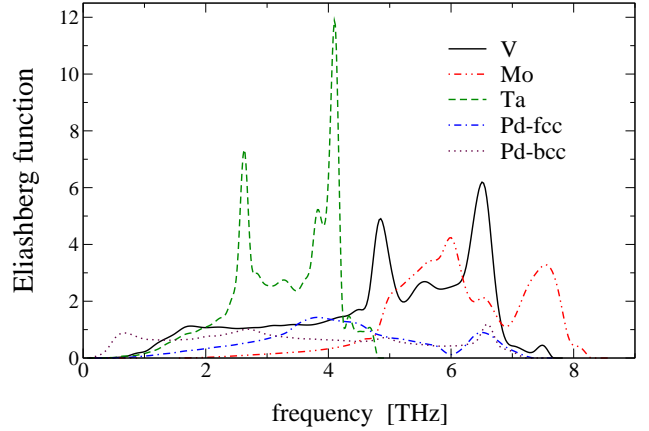}} \caption{\label{a2F}
The Eliashberg functions of V, Mo, Ta in bcc structure and Pd in both bcc and fcc structures.}
\end{figure}
\begin{figure} \epsfxsize=8cm \centerline{
\includegraphics[scale=0.30,angle=-90.0]{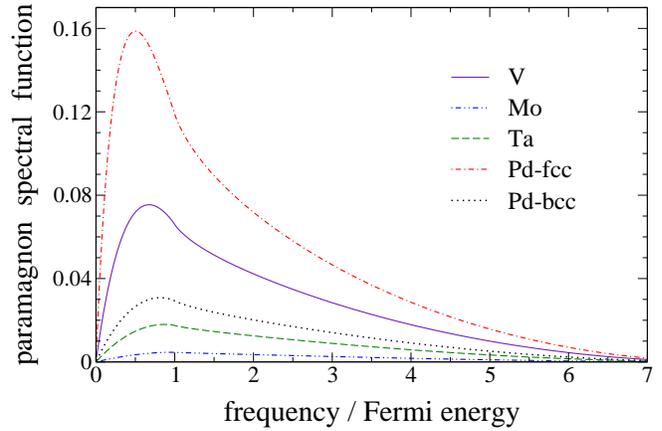}} \caption{\label{Pwpic}
The paramagnon spectral functions of V, Mo, Ta in bcc structure and Pd in both bcc and fcc structures.}
\end{figure}

\begin{table*}
\begin{tabular}{lcccccccccccc}
\hline \hline \\
 &  & & & & && \multicolumn{2}{c}{Eliashberg} & \multicolumn{2}{c}{SCDFT} &&  \\
system & $\;\;$ N($\varepsilon_{F}$) & $I_{ex}$ & $\;\;$ $\lambda^{ph}$
&  $\lambda^{sf}$  & $\mu^{\ast}$ &$\;$&
$T_{c}^{ep}$ & $T_{c}^{epsf}$ &
$T_{c}^{ep}$ & $T_{c}^{epsf}$ &
$\sim T_{c}^{exp}$ & "error"   \\[0.2cm]
\hline \\[0.05cm]
V   bcc & 24.98 (26.14$^{a}$) & 0.0218$^{b}$ &
 0.91 (1.19$^{a}$) & 0.430$^{c}$  & 0.212 &&
  9.0  &  5.9 & 16.1 & 7.4  & 5.38 & 38   \\[0.05cm]
Mo  bcc & $\;\;$8.81 $\;$(8.34$^{a}$) & 0.0184$^{b}$ &
 0.47 (0.42$^{a}$)  & 0.024$^{c}$   & 0.198 &&
  0.8  & 0.7  &  1.5 & 1.4  & 0.92 &  52    \\[0.05cm]
Ta  bcc & 18.60 (18.38$^{a}$) & 0.0162$^{b}$ &
  0.97 (0.86$^{a}$) & 0.096$^{c}$   & 0.209 &&
  8.7  &  8.1  &  5.9  & 4.6  & 4.48 & 3   \\[0.1cm]
Pd  fcc & 30.68 (34.14$^{a}$) & 0.0230$^{b}$ &
  0.35 (0.35$^{a}$)  & 0.972$^{c}$   & 0.213 &&
  0.01  &  \--  & --  & --   & \--  &  \--     \\[0.05cm]
Pd  bcc & 16.60 (18.49$^{b}$) & 0.0229$^{b}$ &
  0.68 (\--\--)   & 0.167$^{c}$  & 0.208 &&
  1.3  &  0.8  & --   & --   & \--  & \--    \\[0.2cm]
\hline \hline \\[-0.1cm]
\multicolumn{13}{l}{$^{a}$ Values from Ref.~\cite{Sav-lambda}.} \\
\multicolumn{13}{l}{$^{b}$ Values from Ref.~\cite{Singalas}.} \\
\multicolumn{13}{l}{$^{c}$ Calculated with $I_{ex}$ from Ref.~\cite{Singalas}.}
\end{tabular}
\caption{ \label{metals} Various parameters such as: the crystal symmetry,
density of states N($\varepsilon_{F}$) per Ry and per both spins,
coupling constants $I_{ex}$  [Ry/both spins],
electron-phonon $\lambda^{ph}$, electron-paramagnon $\lambda^{sf}$,
and $T_{c}$ [K] calculated from the Eliashberg theory and the SCDFT
with the Coulomb and phonon interactions only (ep) and with spin fluctuations (epsf),
the experimental $T_{c}^{exp}$ (from Ref.~\cite{expTc}),
and the "error" defined as $(T_{c}^{epsf} - T_{c}^{exp})/T_{c}^{exp}$ [$\%$] with $T_{c}^{epsf}$
calculated within the SCDFT. }
\end{table*}

In TABLE~\ref{metals}, we report the critical temperatures and parameters
which enter the gap equation calculated by means of the Eliashberg theory and the SCDFT
for a few simple metals: vanadium, molibdenium and tantallum in bcc lattice structure
and palladium in fcc and bcc structures.
Our calculated densities of states, $N(\varepsilon_F)$,
and electron-phonon coupling constants, $\lambda^{ph}$, are in a good agreement
with previous calculations by Savrasov {\em et al.} \cite{Sav-lambda}.
The Eliashberg functions calculated within the DFPT are presented in FIG.~\ref{a2F}.
The Coulomb parameter, $\mu^{\ast}$, was obtained from the Bennemen-Garland
formula \cite{mu,our-Nb}, which employs the density of states.
The spin exchange constant, $I_{ex}$, taken from Ref.~\cite{Singalas},
has been used to obtain the paramagnon spectral function, $P(\omega)$, which we draw in
FIG.~\ref{Pwpic}.

As for the critical temperatures,
for tantallum, the SCDFT result is in a very small relative error, defined in TABLE~\ref{metals},
of 3$\%$  with respect to the experimental data \cite{expTc}.
While, the Eliashberg result with spin fluctuations included is in the error of 81$\%$.
For molibdenium, $T_{c}$ from the Eliashberg gap equation is smaller than the experimental one,
even without the paramagnon effect. But the absolute error of all calculated temperatures for Mo
is smaller than 1~K.
Palladium in both structures fcc and bcc is nonsuperconducting and the SCDFT reproduces well
this result. In contrast to the SCDFT result, from the Eliashberg theory
we obtained superconductivity for Pd in the bcc structure with a very small $T_{c}$.

Usually, the critical temperatures from the SCDFT are lower than temperatures from the Eliashberg
theory. In some cases, however, the SCDFT temperatures are higher.
This situation is for vanadium and molibdenium.
Especially for vanadium, $T_{c}$ from the SCDFT gap equation is about 2~K higher
than the experimental data \cite{expTc}, even after inclusion of spin fluctuations.
This fact may indicate that, either the spin exchange constant, $I_{ex}$, was underestimated,
or  a contribution of the asymmetric part of the phononic term in the SCDFT
 gap equation is quite large. As we know from results reported in Refs.~\cite{Martin-OGK,our-Nb},
if we neglect the asymmetric part in the electron-phonon-coupling matrix elements by taking
the $\alpha^{2}F(\omega)$ avaraged at the Fermi level, the critical temperatures are higher
(see the discussion in Section~\ref{implement}). The last approximation, however, has to be done
if we do not evaluate formulas with the $g_{{\bf k,k+q}}$ elements explicitely.

\begin{table*}
\begin{tabular}{cccccccccccc}
\hline \hline \\
 &  & & & & &   \multicolumn{2}{c}{Eliashberg} & \multicolumn{2}{c}{SCDFT}  & &  \\
 $p$ & N($\varepsilon_{F})$ & $I_{ex}$ &
$\lambda^{ph}$  &  $\lambda^{sf}$ && 
$T_{c}^{ep}$ & $T_{c}^{epsf}$ & 
$T_{c}^{ep}$ & $T_{c}^{epsf}$ & 
$\sim T_{c}^{exp}$ &  "error"  \\[0.2cm]
\hline \\[0.05cm]
 -16.59 & 22.82  & 0.0211 & 1.91  & 0.28  &&  20.3 & 16.7 &  14.4 & 6.2  &  \--  & \--  \\[0.05cm]
 -9.45  & 21.60  & 0.0213 & 1.60  & 0.25  &&  19.5 & 15.5 &  13.2 & 6.4  &  \--  & \--  \\[0.1cm]
 -0.63  & 20.24  & 0.0217 & 1.41  & 0.22  &&  18.8 & 14.7 &  12.9 & 7.2  &  9.2  & -22  \\[0.05cm]
  9.98  & 19.38  & 0.0204 & 1.65  & 0.17  &&  19.6 & 15.8 &  13.4 & 9.8  &  10.0 & -2    \\[0.05cm]
  22.89 & 18.32  & 0.0189 & 1.47  & 0.13  &&  19.4 & 16.0 &  13.2 & 11.3 &  9.8  & 15    \\[0.05cm]
  38.79 & 17.10  & 0.0228 & 1.29  & 0.16  &&  18.4 & 14.1 &  12.0 & 10.1 &  9.7  &  4    \\[0.05cm]
  56.73 & 15.42  & 0.0292 & 1.10  & 0.23  &&  16.1 & 10.7 &  10.1 & 8.4  &  9.5  & -12  \\[0.05cm]
  78.37 & 13.10  & 0.0347 & 0.86  & 0.24  &&  13.7 &  7.3 &  8.2  & 7.9  &  8.8  & -10  \\[0.2cm]
\hline \hline
\end{tabular}
\caption{ \label{Tab-Nb} Results for Nb; applied pressure $p$ [GPa],
density of states  N($\varepsilon_{F})$ per Ry and per both spins,
spin exchange integral $I_{ex}$ [Ry/both spins] (from Ref.~\cite{our-Nb}),
coupling constants: electron-phonon $\lambda^{ph}$, electron-paramagnon $\lambda^{sf}$,
and $T_{c}$ [K] calculated from the Eliashberg theory (with $\mu^{\ast}$=0.21) and SCDFT
with Coulomb and phonon interactions only (ep) and with spin fluctuations (epsf).
The experimental $T_{c}^{exp}$ has been estimated from the picture given
in Ref.~\cite{Struzhkin}.
Last column shows the "error" of the SCDFT calculations for $T_{c}^{epsf}$ defined in TABLE~\ref{metals}.}
\end{table*}

In general, the critical temperatures obtained from the SCDFT are in a good agreement with
the measured temperatures \cite{expTc}, and the effect of  paramagnons improves the result
considerable for many simple metals.

\subsection{Niobium under pressure}

In TABLE~\ref{Tab-Nb}, we present critical temperatures and parameters of the gap equation 
for niobium at eight pressures in the range from -17~GPa up to 80~GPa. 
The spin exchange constants, $I_{ex}$, 
have been calculated from first principles in Ref.~\cite{our-Nb}, and 
the electron-phonon and electron-magnon spectral functions for Nb have been presented also 
in that work.

Here, we complete our previous results by reporting the effect of paramagnons on $T_{c}$ calculated 
from the SCDFT. After the inclusion of spin fluctuations, the critical temperatures obtained from
the SCDFT are closer to the experimental $T_{c}$'s for pressures in the range of 0\--40~GPa, 
{\em i.e.} pressures between two anomalies measured by Struzhkin {\em et al.} \cite{Struzhkin}. 
The dependence of the measured critical temperature as a function of pressure 
is no longer reproduced by our calculations when we take into account paramagnons. 
At ambient pressure and for higher pressures, paramagnons seem to make 
the theoretical result worse.
The above effect, could be explained by making the observation that, in every case where 
the exchange constant $I_{ex}$ is large, the theoretical temperature underestimates the 
measured temperature, and vice versa, for the smallest $I_{ex}$ the critical
temperature obtained from the SCDFT is the highest and the error is positive.         

Concluding this Subsection, the implementation of paramagnons to the SCDFT generally makes
calculated critical temperatures closer to the experimental ones. 
But our calculated exchange constants, $I_{ex}$, are not sufficiently accurate. 
This fact gives a direction for the future development.

%% file: summary.tex
\section{Summary}

In the present work, we included the transverse spin fluctuations to the density 
functional theory for superconductors.
The SCDFT is presented from its foundations, through the decoupling  approximation,
the gap equation and details of implementation.
We assumed singlet and $s$-wave pairing potential; The extension to triplet
superconductors could be done following the work by Capelle {\em et al.} 
\cite{KC,triplet}. The electron-phonon couplings and the electron-paramagnon couplings were 
averaged at the Fermi energy, therefore the asymmetric part of the functional with respect to 
the electron-phonon matrix  elements and to the spin-exchange interaction constants were ommited.  
Through the whole work, we kept the notation to be consistent with Parks \cite{Parks,Kazumi} 
and Vonsovsky \cite{Vonsovsky}.  

Paramagnons and phonons in the superconducting state were assumed to be the same like
in the normal state. The Eliashberg spectral function has been calculated within the
density functional perturbation theory and it is fully material specific. 
Paramagnons, in contrast, have been obtained from the random phase
approximation for the homogeneous electron gas and only the spin exchange constants were 
calculated from the electronic structure. 

We reported the critical temperatures obtained from the SCDFT and 
the Eliashberg linearized gap equation with and without spin fluctuations   
for a few simple metals: V, Mo, Ta, Pd at ambient pressure and Nb at
several pressures up to 80~GPa. 
Some discrepancies between the temperatures calculated from the SCDFT and the measured temperatures
are due to the fact that it is quite difficult to obtain the accurate spin-exchange constants and/or
to the fact that the spectral functions have been averaged at the Fermi level.
Netherveless, the results show that inclusion of paramagnons improves the critical temperatures obtained  
from both methods, the SCDFT and the Eliashberg theory. The critical temperatures obtained 
from the parameter-free SCDFT are in most cases closer to the experimental data 
than the results obtained from the Eliashberg theory.